# Optimal Ultra-wide Spatial-Spectral Windows for Hyperentangled Two-photon Emission

S. F. Hegazy[1], J. El-Azab[1], Y. A. Badr[1], and S. S. A. Obayya[2*]

[1] National Institute of Laser Enhanced Sciences, Cairo University, 12613 Giza, Egypt
[2] Centre for Photonics and Smart Materials, Zewail City of Science and Technology, 12588 Giza, Egypt
[*]corresponding author: sobayya@zewailcity.edu.eg

**Abstract**- While being optimally compensated for spatial phase variations, the two-photon state produced by the two-crystal emission exhibits spatial and spectral decoherence off the central emission modes. In this paper, we present an experimentally convenient method to optimize the ultra-wide spatial and spectral windows; allowing the minimum spatial-spectral decoherence for a required two-photon flux.

The creation of hyperentangled photons entails two-photon emission over relatively wide extent in frequency and transverse space within which the photons are simultaneously entangled in energy, spatial mode and polarization [1-3]. Because the creation process runs in nonlinear domain(s) which is always dispersive and birefringent, the output two-photon state undergoes loss of relative-phase coherence over frequency and space. This offers the vital role of spatial-spectral phase compensation using birefringent element [2, 4] so as to restore partially the state coherence in the two degrees of freedom. After compensation, the two-photon state emerges with much better phase flatness allowing collection over wider spatial and spectral ranges. However, over ultra-wide windows, the spatial and spectral modes away from the central *truly* compensated modes suffer strong phase variations appears to dominate the scene. In this paper, we present an experimentally convenient approach to optimizing the ranges of the ultra-wide spatial and spectral filters so as to diminish decoherence of the two-photon state at fixed photon-pair flux. Consider a paraxial pump field linearly polarized on 45 deg. and described as a sum of monochromatic plane waves $E_p(r,t) = \iint dw_p dq_p A_p(w_p, q_p) \exp i(k_p \cdot r - w_p t)$. As depicted in Fig. 1(a), this pump field illuminates two contiguous identical type-I crystals whose principal planes are orthogonal to each other, the two-photon state (excluding vacuum) can be expressed to first order as

$$|\psi\rangle = \int dw_s \, dw_i dq_s \, dq_i \; \chi^{(2)}(w_s, w_i; w_s + w_i) \, A_p(w_s + w_i, q_s + q_i) \, d \operatorname{sinc}\left(\frac{\Delta\kappa d}{2}\right)$$
$$\times \left\{|H\rangle_{q_s,w_s}|H\rangle_{q_i,w_i} + \exp i[\Phi_o + \Phi_{\text{DC}}(q_s, q_i, w_s, w_i)]\, |V\rangle_{q_s,w_s}|V\rangle_{q_i,w_i}\right\} \quad (1)$$

where $\chi^{(2)}$ is the second-order effective susceptibility, $(q_s, q_i)$ and $(w_s, w_i)$ are the transverse wavevectors and angular frequencies of the signal and idler photons, respectively, $\Phi_o$ is the initial phase difference between the pump components, $\Phi_{\text{DC}}$ is the relative phase accumulated in the SPDC process, $d$ is the crystal length, and $\Delta\kappa$ is the longitudinal wavevector mismatch. and $A_p(w_p, q_p)$ is assumed to be a real function centered on $w_p = w_p^0$ and $q_p = 0$. Assuming that the effective susceptibility is slowly varying within that spectral window and the spatial-spectral function $A_p(w_s + w_i, q_s + q_i)$ is very narrow relative to the remainder of the integrand in (1), the probability of the photon pair to be detected along a mode in the momentum-frequency space $(q_s) \otimes (w_s)$ is then

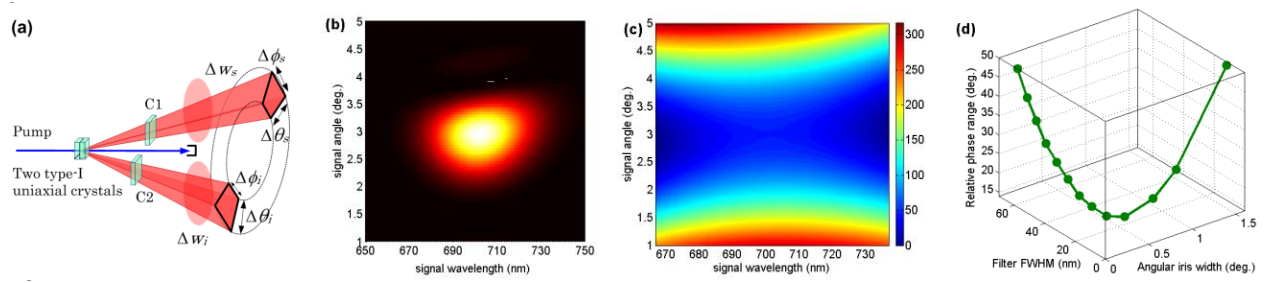

Fig. 1 (a) hyperentangled photon source equipped with optimal compensation crystals and ultra-wide iris and spectral filters. (b) spatial-spectral two-photon probability and (c) relative phase map calculated behind two compensation elements acting by the optimal compensation. The pump laser is taken at 351.1 nm and the SPDC and compensation crystals are BBO crystals with $d = 0.59$ mm and cut angle 33.9° producing degenerate emission centered at ~3° opening angle. The spectral filters are assumed to have a normalized Gaussian profile of 70-nm FWHM. (d) Optimization of residual relative-phase over different arrangements of ultra-wide spatial and spectral windows realizing the fixed two-photon flux.

$$P(q_s, w_s) \propto \left| G(w_s - w_d^0) \ G(w_p^0 - w_s - w_d^0) \ \text{sinc}\left(\frac{\Delta\kappa(q_s, w_s) \ d}{2}\right) \right|^2 \quad (2)$$

behind two filters of spectral profile $G(w)$ centered at the degenerate frequency $w_d^0$. One should notice that since the SPDC crystals are thin and the two-photon emission is at small angles, both the probability $P(q_s, w_s)$ and the phase map $\Phi_{DC}(q_s, w_s)$ depend negligibly on the azimuthal angle [5].

A major request by an experimentalist is thereby to obtain the FWHM of the spectral filter as well as the width of the iris that allow a required rate of coincidence counts while optimally minimize the phase map difference. Given the spatial-spectral two-photon probability in (2) [see Fig. 1.(b)], one can determine an array of arrangements of ultra-wide spatial and spectral windows realizing fixed two-photon flux. Now, recalling the compensated relative-phase map obtained in ref. [4] [see Fig. 1.(c)], each of these fixed-flux arrangements leads to a particular range of phase map variations. The points in Fig 1.d therefore represent the phase-map range corresponding to the variant fixed-flux arrangements. For the optical setup parameters in Fig. 1, a 30-nm interference filter and ~0.5-deg acceptance iris are found offering the minimum phase map variations compared with the other arrangements of the same pair flux.

**Acknowledgements**, This work was supported by the Information Technology Industry Development Agency (ITIDA-ITAC), ministry of communication, Egypt.